\documentclass[fleqn,usenatbib]{mnras}

\usepackage{newtxtext,newtxmath}

\usepackage[T1]{fontenc}

\DeclareRobustCommand{\VAN}[3]{#2}
\let\VANthebibliography\thebibliography
\def\thebibliography{\DeclareRobustCommand{\VAN}[3]{##3}\VANthebibliography}

\usepackage{graphicx}	
\usepackage{amsmath}	

\usepackage{subfigure}
\usepackage{soul}

\title[Kepler Twins]{A List of 49 New Stellar Twins from the Kepler Catalog of Eclipsing Binary Stars}

\author[Y\"{u}cel et al.]{
G\"{o}khan Y\"{u}cel,$^{1}$\thanks{E-mail: gokhanyucel@akdeniz.edu.tr}
Volkan Bak{\i}\c{s}$^{1}$
\\
$^{1}$Akdeniz University, Department of Space Sciences and Technologies, Dumlup{\i}nar Blv., Kamp\"{u}s, 07058, Antalya, TR\\
}

\date{Accepted XXX. Received YYY; in original form ZZZ}

\pubyear{2022}

\begin{document}
\label{firstpage}
\pagerange{\pageref{firstpage}--\pageref{lastpage}}
\maketitle

\begin{abstract}
49 new eclipsing twin binary candidates are identified and analyzed based on Kepler eclipsing binary light curves. Their colours and spectral types are calculated according to our classification. A comparison of the spectral type distribution of eclipsing twin binary systems showed that F-type twins dominate among others, which agrees well with recent studies. The distance of eclipsing twin binaries from the galactic plane shows that F and G-type twins can be seen at any distance from the galactic plane and most of the known eclipsing binary twins are located within 200 pc of the galactic plane, which could be interpreted as these systems are members of thin disk population. As a case study, a twin binary system selected from our updated list of twins, V396~Gem, has been analyzed with spectroscopic and Kepler data. As a result, we have derived the physical parameters of the components of  V396~Gem as $M_{1,2}(M_\odot)= 1.814\pm0.114$, $1.797\pm0.114$; $R_{1,2}(R_\odot)= 2.655\pm0.078$, $2.659\pm0.090$; $T_{\mathrm{eff}_{1,2}}(K)=7000\pm100$, $6978\pm100$; $[M/H]=0.11\pm0.03$. We have calculated the evolutionary status of the components by using MESA. Accurately derived physical parameters of the components of V396~Gem have allowed us to determine the age of the system as 1.168$\pm$0.149 Byrs.
\end{abstract}

\begin{keywords}
Surveys -- binaries: eclipsing -- techniques: spectroscopic -- stars: fundamental parameters 
\end{keywords}



\section{Introduction}

The first study on twin binaries with known mass ratio ($q$, $M_2$/$M_1$) was achieved by \cite{lucy1979}. In their study, they listed binaries according to their mass ratio, and found that there was a peak at $q$ $\approx$ 0.97. Similar studies have been conducted by various researchers on specific binary systems [giants and dwarfs \citep{Jaschek1972}, Solar-type stars \citep{abt1976}, early B-type stars \citep{abt1978}, double-lined spectral binary stars \citep{Kraicheva1978}, late B-type stars \citep{Stani1979,Popova1982}, O-type stars \citep{Garmany1980,Fofi1983}, main sequence binary systems \citep{Griffin1985,Halbwachs1987}]. All of these studies have a common ground that the mass ratio distribution of SB2 (double-lined spectroscopic binary) type systems shows a net peak around $q$ $\approx$ 1 and a shallow peak around $q$ $\approx$ 0.2. 

\cite{Halbwachs2003} has studied binary systems with primary components with F-G-K spectral type and with periods of up to 10 years. They concluded that there are two broad peaks on $q$ $\approx$ 0.2 and $q$ $\approx$ 0.7, and a sharp peak at $q$ $\approx$ 0.8 but there are no spectral type differences among the mass ratio distribution. Later on, \cite{Simon2009} conducted a research, based on the 9th Catalogue of Spectroscopic Binaries (SB9), and discovered that most of the twin binaries are of F-G-K spectral types.

But all these studies have been conducted by already analysed and identified twin binary systems and none of them has focused on finding and analyzing new twin systems until recent years. Recently, \cite{bakis2020} have investigated the All Sky Automated Survey \citep[ASAS,][]{pojmanski} for twin binaries that have $\delta$ > -30 $\deg$, and identified 68 new twin binary systems. They observed the systems by photometric and spectroscopic methods, and determined the spectral type of the systems. To achieve this, they found absolute colours of systems via help of photometric data. Also via the help of spectroscopic data, they built a grid of atmosphere models and determined the spectral types of systems by modelling observed spectra with comparing synthetic spectra. As a result, they showed that twins are, indeed, mostly in F-G-K spectral type, but there is a clear tendency towards F-type.

Is that really the case? Are eclipsing twin binaries favour F-spectral type among late type spectra in every region of the sky? With the motivation to answer these questions, we enlarged the twin sample in the present paper with the Kepler data. The Kepler space telescope was operated between 2009--2013 (Kepler original mission, \cite{kepler}) and 2014--2018 (K2 mission, \cite{Howell}) to find Earth-like planets in habitable zones near Solar-like stars. Even though its main object is to find exoplanets, the data generated by the Kepler satellite also help researchers to find and analyze binary stars \citep[e.g.][]{helminiak2017hides,matson}. More importantly, since Kepler was looking for relatively late spectral type stars, it is a perfect tool for looking up to investigate twin binaries among late spectral types. Also, since Kepler looks at a specific region of the sky, it might help us to answer the question whether there is a selection effect due to the location of the stars. \cite{Zhang2017} also studied eclipsing twin binary systems in Kepler data. They selected candidates based on the depth and width ratio of eclipses in the light curves of the systems. According to their classification, if the differences between depth and width ratio on light curve eclipses is less than 2\%, the system is suitable for a twin candidate. By this methodology, they found 28 new twin candidate systems. They also found the mass ratio of selected candidates via $q$-search method and indicated that two of the selected candidates were suited for their "twin criteria" ($q$ > 0.95). \cite{Zhang2017} analyzed the Kepler data with a limited amount of low resolution spectroscopic data, and determined the fundamental parameters of the systems. They did not perform any analysis for the remaining systems.

In the present study, we have selected candidate systems from the Kepler Eclipsing Binary Catalog\footnote{\url{http://keplerebs.villanova.edu/}} \citep[hereafter KEBC,][]{Prsa,Slawson,Mati,Kirk} (\S\ref{sec:kep_data}). Before examining the data, we have selected suitable candidates from the catalog and determined the twin candidates with a classification criteria, which is more stringent (see \S\ref{sec:obs}) than given by \cite{Zhang2017}. Kepler data of the candidates have been analyzed and the final list of eclipsing twin binaries was obtained according to our classification criteria. Photometric observations were carried out and spectral types of the candidates were obtained in (\S\ref{sec:extinction}). In \S\ref{sec:case_study}, fundamental parameters of a secure twin, V396~Gem, are presented as a case study based on spectroscopic and photometric analysis. Finally, in \S\ref{sec:discussion}, our findings are discussed.

\section{Methodology}
\label{sec:methodology}

Our main methodology for finding eclipsing twin binary systems is based on \cite{bakis2020}, where the details are given in the following subsections.

\subsection{Kepler data, pre-selection of candidates and constructing the final list of twins}\label{sec:kep_data}

The details of the Kepler spacecraft and photometer have been discussed very well by \cite{kepler}, \cite{Koch}, \cite{Caldwell}, and others, therefore, details could be found in those studies. We have selected our initial candidates from the KEBC (last updated at August 8, 2019), which is a catalog providing the basic properties of listed eclipsing binary systems, such as Kepler magnitude, equatorial and galactic coordinates, orbital period, mid-eclipse time and components' effective temperature. One of the useful features of KEBC is that it also provides a morphological index of the systems, which classifies binary systems as detached, semi-detached or contact binary. This feature has saved us a lot of time in the selection of twin candidates. KEBC, additionally, provides depths and widths of eclipses of the light curves, based on polyfit \citep[][]{Prsa2008}, which helped us to build our initial twin candidate list according to our classification.

\begin{figure}
    \begin{center}
        \includegraphics[width=0.45\textwidth]{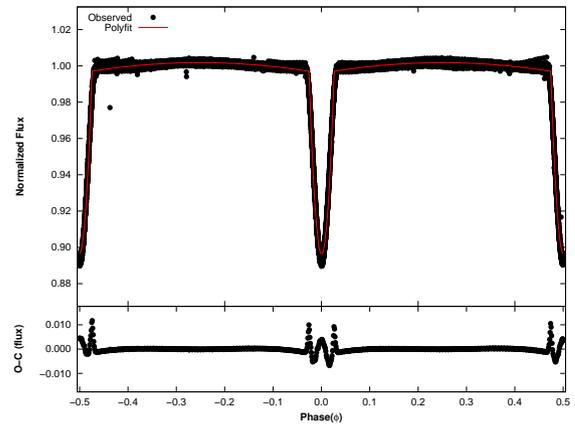}
\caption{Original Kepler data (filled points) and best fitting polyfit (red solid line) for eclipsing binary KIC 4365461.}
    \label{fig:polyfit}
    \end{center}
\end{figure}

Even though polyfit, as reliable as, helped us to select twin candidates to analyze, we have selected initial candidates as systems with an eclipse depth difference of 0.006 mag between their primary and secondary minima. Selecting this criteria is based on the accuracy of polyfit near mid-eclipses of binary light curves. As the polyfit uses chain of \textit{n}th order polynomials, it gives efficient accuracy for eclipse depths \citep[see details of polyfit in][]{Prsa2008}. An example is shown in Fig.\ref{fig:polyfit} with an RMS of residuals of 0.004 in the phase range $\Delta\phi=$--0.01-+0.01. The value of eclipse depth difference was specifically selected as small as possible, considering the accuracy of Kepler photometry as 0.001 mag. Albeit this selection criteria would cause us to overlook some twin systems, such as systems with evolved components (see \S\ref{sec:discussion}), it helped us to find twins more precisely. In the selection criteria, the difference between the widths of primary and secondary minima has not been considered since we planned to analyze the individual light curves of each twin candidate in detail. Among the 2950 eclipsing binaries listed in KEBC, it is found that 88 circular orbit binary ($\sim$3 per cent of all eclipsing binaries) systems are suitable for our selection criteria. As for systems with eccentric orbits, this selection criterion is useless as the eclipse depths of systems with eccentric orbits may change independently of the temperature ratio of the components since the components change their separation during the orbital motion, which causes a change in the eclipsed area and, hence, the eclipse depth. Therefore, it is necessary to solve the light curves of eccentric orbit candidates to determine the twins. Recently, \cite{Kjurk2017} published a study to identify systems with an eccentric orbit in KEBC. In this study, \cite{Kjurk2017} determined a total of 529 eccentric orbit systems with reliable orbital parameters (eccentricity ($e$) and longitude of periastron ($w$)). However, as they stated, light curve parameters they obtained such as relative radii and effective temperatures of the components are not as reliable as the eccentricity and longitude of periastron. Therefore, in order to determine an eccentric twin from their list, a light curve analysis of a candidate is necessary.Thus, we selected 47 eccentric twin candidates for further analysis using the ratio of relative radii and temperatures of the components which are provided by \cite{Kjurk2017}.  The initial twin candidate list and their basic properties (circular and eccentric orbit) are presented in Table~\ref{tab:initial}.

\begin{table*}
  \caption{Basic information of initial twin candidates. The full table is available online.}
    \resizebox{\textwidth}{!}{\begin{tabular}{ccccccccccc}
    \hline
    Order & KIC / EPIC & Period & T$_0$ & RA & DEC & K$_{mag}$ & \multicolumn{2}{c}{Eclipse Depth} & \multicolumn{2}{c}{Eclipse Width} \\
    & (ID) & (days) & (BJD-2400000) & (deg) & (deg) & (mag) & pri. & sec. & pri. & sec. \\
    \hline
    1     & 2305543 & 1.3622735 & 55003.400909 & 292.0269 & 37.6007 & 12.545 & 0.1078 & 0.1052 & 0.0659 & 0.0687 \\
    2     & 2306740 & 10.3069870 & 54966.42521 & 292.2698 & 37.6982 & 13.545 & 0.3153 & 0.2950 & 0.03860 & 0.0209 \\
    3     & 3248019 & 2.6682217 & 55098.773000 & 294.5968 & 38.3701 & 15.387 & 0.0475 & 0.0462 & 0.0417 & 0.0417 \\
    4     & 3654950 & 8.1347246 & 55003.927327 & 293.8329 & 38.7363 & 15.858 & 0.0559 & 0.0585 & 0.0307 & 0.0245 \\
    ... & ... & ... & ... & ... & ... & ... & ... & ... & ... & ... \\
    \hline \label{tab:initial}
    \end{tabular}}%
\end{table*}%

The light curve analysis of the candidates listed in Table~\ref{tab:initial} were performed with the Physics of Eclipsing Binary (\textsc{PHOEBE}) \citep{Prsa2005} user interface, which uses W-D code \citep[][]{1971ApJ...166..605W,1979ApJ...234.1054W,1990ApJ...356..613W,2008ApJ...672..575W,2010ApJ...723.1469W,2014ApJ...780..151W}. Although we focused on only temperature ratios in the pre-selection stage, we included the ratio of relative radius of the components in our selection criteria when creating our final list. Therefore, basically, if the temperature ratio and ratio of relative radius of the components are less than 5 per cent, that system was included in the final list.

The final list of twins includes 34 ($\sim$1 per cent of all binaries in KEBC) circular orbit and 15 ($\sim$3 per cent of all eccentric binaries in KEBC) eccentric orbit twin binary systems, which are given in Table \ref{tab:observed_circ} and Table \ref{tab:eccentric}, respectively. One system in our list, KIC 5738698, has already been proven by spectroscopic analysis that it is a twin-binary system by \cite{matson}. Also, 5 twin candidate systems, KIC 4157488, KIC 4365461, KIC 4826439, KIC 8330575, and KIC 10480952, which are given by \cite{Zhang2017} are also on our final list.

\subsection{Observations}
\label{sec:obs}

Photometric observations of circular orbit twin systems have been performed using two telescopes, UBT60\footnote{\url{http://fen.akdeniz.edu.tr/uzay-bilimleri-ve-teknolojileri/arastirma/teleskoplar/}} and T100\footnote{\url{https://tug.tubitak.gov.tr/en/teleskoplar/t100-telescope}}, which are equipped with CCDs, Alta U47 and SI1100, respectively. The telescopes are located at 2500-m altitude of T\"{U}B{\.{I}}TAK National Observatory (TUG) observing site in Bak{\i}rl{\i}tepe, Antalya, Turkey. Photometric observations were made in years between 2018-2020 using standard Johnson $UBVRI$ filters. Most of the program stars are relatively faint for our telescopes, therefore, $U$-band observations were performed for only relatively bright systems (19 out of 34) for more reliable measurements. Observed systems were transformed into standard magnitudes using the transformation equations  given by \cite{bakis2020}. The standard stars used for transformations were selected from \cite{Landolt}. Our final list of twin systems with a circular orbit, observed colours, measured $V$-band magnitude, and light contribution ratio of the systems are given in Table~\ref{tab:observed_circ}. Twins with an eccentric orbit were not observed photometrically. Their spectral types were determined using their temperatures given by \cite{Pinson2012}. Therefore, observed colours of the eccentric orbit twins are not listed in Table~\ref{tab:observed_circ}

\begin{table*}
  \centering
  \caption{Final list and observed quantities of twin systems with a circular orbit. The full table is available online.}
\resizebox{\textwidth}{!}{
\begin{tabular}{ccccccccccc}
\hline
No    & KIC / EPIC  & r$_1$/r$_2$ & T$_1$/T$_2$ & $(U-B)$   & $(B-V)$   & $(V-R)$   & $(V-I)$  & $(R-I)$   & V$_{{\mathrm obs}}$ & $l_1/l_{\rm total}$ \\
& ID & & & (mag) & (mag) & (mag) & (mag) & (mag) & (mag) & \\
\hline
    1     & 3654950 & 0.998 & 1.014 &   --  & 1.035$\pm$0.186 & 0.669$\pm$0.061 & 0.516$\pm$0.053 & 1.186$\pm$0.114 & 16.545$\pm$0.166 & 0.517 \\
    2     & 4157488 & 0.997 & 0.999 &   --  & 0.685$\pm$0.053 & 0.377$\pm$0.001 & 0.439$\pm$0.034 & 0.816$\pm$0.035 & 14.093$\pm$0.130 & 0.498 \\
    3     & 4365461 & 0.999 & 1.000 &   --  & 0.582$\pm$0.071 & 0.348$\pm$0.006 & 0.359$\pm$0.002 & 0.707$\pm$0.008 & 13.501$\pm$0.157 & 0.498 \\
    ... & ... & ... & ... & ... & ... & ... & ... & ... & ... & ... \\
\hline
\label{tab:observed_circ}
\end{tabular}}
\end{table*}

\begin{table*}
  \centering
  \caption{Final list of twin systems with eccentric orbit. The full table is available online.}
    \begin{tabular}{ccccccc}
    \hline
    No    & KIC / EPIC  & r$_1$/r$_2$ & T$_1$/T$_2$ & $e$ & $w$ & $l_1$/$l_{\rm total}$ \\
    & ID & & & & (degree) & \\
    \hline
    1     & 3865298 & 0.966 & 0.994 & 0.233 & 171   & 0.477 \\
    2     & 4375101 & 1.012 & 0.967 & 0.230 & 24    & 0.478 \\
    3     & 4937143 & 1.007 & 1.002 & 0.211 & 1     & 0.501 \\
    ... & ... & ... & ... & ... & ... & ... \\
        \hline \label{tab:eccentric}
    \end{tabular}%
\end{table*}%

\subsection{Determining Spectral Types and Interstellar Extinction} \label{sec:extinction}

Eclipsing twin binary systems consist of two identical components, which have identical colours independent of the orbital phase. Therefore, observed photometric color of twins obtained at any orbital phase may be used as an indicator of its spectral type. However, the interstellar reddening must be determined to have intrinsic colours of the systems. In order to obtain the unreddened colours, we have built colour-colour (CC) diagrams based on unreddened colours of $UBVRI$ data for spectral classes from O3 to M6 given by \cite{Pecaut}\footnote{\url{http://www.pas.rochester.edu/~emamajek/EEM_dwarf_UBVIJHK_colors_Teff.txt}}. After that, we located the systems on the CC diagrams. This procedure enabled us to determine color excess of an eclipsing twin binary system by using the galactic extinction law given by \cite{Fitz2019}. Reddening ratios were adopted from \cite{Fitz2019} as 0.793, 0.751, and 1.572 for $E(U-B)/E(B-V)$, $E(V-R)/E(B-V)$, and $E(V-I)/E(B-V)$, respectively. Thus, color excesses of the systems have been found, and true colours of the systems, which helps us to determine the spectral type of that system, have been obtained. A list of spectral types, unreddened $(B-V)_0$ colours, temperatures and $E(B-V)$ color excesses are given in Table~\ref{tab:Results}.

\begin{table}
  \centering
  \caption{Determined spectral types from the analyzed photometric data and temperatures, which calculated from \protect\cite{Eker2020}. The full table is available online.}
  \label{tab:Results}
    \resizebox{0.48\textwidth}{!}{\begin{tabular}{lcccccc}
    \hline
    No    & Name & Sp. Type & $(B-V)_0$ & $T_\mathrm{eff}$ & $E(B-V)$ & d$_{GAIA}$ \\
   \hline
    1     & KIC 3654950  &  G4-G6  &  0.670-0.700  &  5499-5392   &  0.355$\pm$0.186  & 2625 \\
    2     & KIC 3865298  &  G9-K0  &  -    & 5295  &  -    & 1341 \\
    3     & KIC 4157488  &  G4-G7  &  0.670-0.710  &  5499-5357   &  0.005$\pm$0.053  & 2227 \\
    4     & KIC 4365461  &  F4-F7  &  0.410-0.510  &  6605-6138   &  0.098$\pm$0.071  & 1419 \\
    ... & ... & ... & ... & ... & ... & ... \\
    \hline
    \end{tabular}}%
\end{table}%

\section{Case Study: V396~Gem}\label{sec:case_study}

Among our list, an F-type eclipsing twin binary, V396~Gem, was selected for a more detailed analysis. Spectroscopic observations were made using the Shelyak Instruments eShel Spectrograph (SIeS), R\,$\sim$\,12000, attached to 0.6-m UBT60 telescope. Observations were made on 7 nights between March-April of 2021. Since V396~Gem is relatively faint ($V\sim10.3$\,mag) for our telescope and spectrograph limits, exposure times were set to the 1200s in a sequence of 6 exposures to achieve a S/N ratio between 50--100 when combined. The spectra were reduced, wavelength calibrated, and continuum normalised by using Image Reduction and Analysis Facility, {\textsc{iraf}}\footnote{{\sc iraf} is provided by National Optical Astronomy Observatories (NOAO) in Tuscon, Arizona, USA \citep[][]{Tody}.}.

\begin{figure}
    \begin{center}
        \includegraphics[width=85mm]{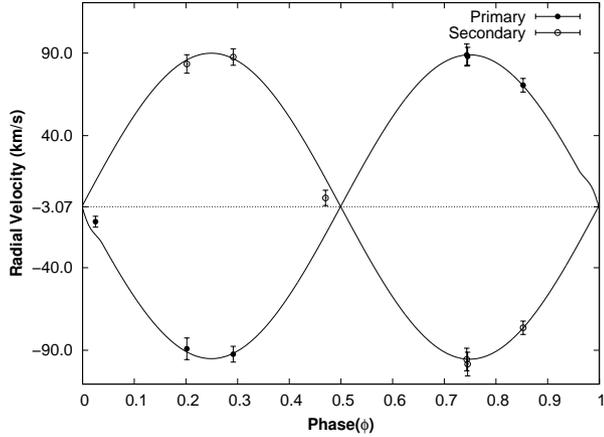}
\caption{Spectroscopic orbit for V396~Gem.}
    \label{fig:RV}
    \end{center}
\end{figure}

\begin{figure}
\begin{center}
\includegraphics[width=85mm]{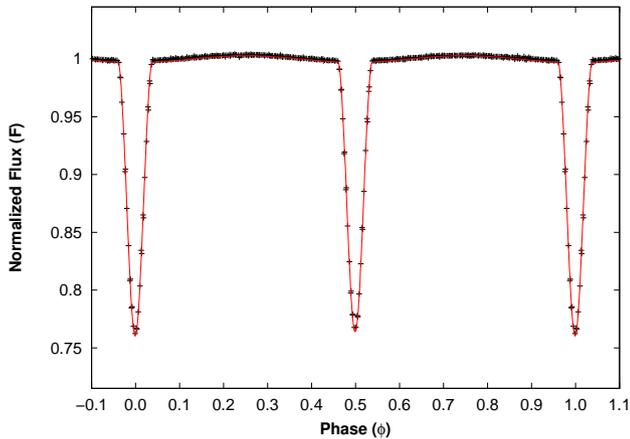}
\caption{Light curve model for V396~Gem.}
\label{fig:LC}
\end{center}
\end{figure}

For determining the spectroscopic orbit, radial velocities of the components of V396~Gem have been obtained by the cross correlation technique using {\sc fxcor} task. Measured radial velocities and their uncertainties are presented in Table~\ref{tab:rv}, and obtained parameters of the spectroscopic orbit of V396~Gem are presented in Table~\ref{tab:orbit}. By using the mass ratio that was determined from spectroscopic orbit, a new light curve model has been obtained using Kepler photometric data of the system. LC model parameters of V396~Gem are given in Table~\ref{tab:LCparameter}. The Spectroscopic orbit and LC model for V396~Gem are shown in Figs.~\ref{fig:RV} and \ref{fig:LC}, respectively.

\begin{table}
\centering
\caption{The RVs of the components of V396~Gem.}
\begin{tabular}{cccccc}
\hline \hline
HJD & Phase  & RV$_{1}$ & $\sigma_{1}$ & RV$_{2}$ & $\sigma_{2}$\\
-2400000 & $\phi$ & (kms$^{-1}$) & (kms$^{-1}$) & (kms$^{-1}$) & (kms$^{-1}$) \\
\hline
59278.42320 & 0.85 & 70.52 & 4.14 & --76.35 & 4.05 \\
59279.37033 & 0.03 & --12.13 & 3.31 & -- & -- \\
59286.33087 & 0.29 & --92.37 & 4.78 & 87.59 & 5.01 \\
59287.31241 & 0.47 & -- & -- & 2.28 & 4.65 \\
59294.31954 & 0.75 & 87.98 & 5.50 & --98.35 & 7.14 \\
59302.32473 & 0.20 & --89.08 & 6.55 & 83.42 & 5.60 \\
59327.28471 & 0.74 & 88.92 & 6.56 & --95.55 & 6.83  \\
\hline \label{tab:rv}
\end{tabular}
\end{table}

\begin{table}
	\small
	\begin{center}
		\caption{Parameters of the spectroscopic orbit of V396~Gem.}
		\begin{tabular}{lc}  \hline \hline
			Parameter                   & Values       \\
			\hline
			$P$(days)                   &   5.4956923 $\pm$ 0.0022831	  \\
			$T_{0}$(HJD-2459277)                & 0.8541 $\pm$ 0.0211 \\
			$K_{1}$(km s$^{-1}$)        & 91.46 $\pm$ 1.71  \\
			$K_{2}$(km s$^{-1}$)        & 92.36 $\pm$ 1.62  \\
			$e$                         & 0.0 (fixed) \\
			$V_{\gamma}$(km s$^{-1}$)   &  -3.47 $\pm$ 1.24  \\
			$q (M_2/M_1)$                         &   0.990 $\pm$ 0.036 \\
			$m_{1}{\mathrm sin}^{3}i (M_{\odot})$ &  1.78 $\pm$ 0.09 \\
			$m_{2}{\mathrm sin}^{3}i (M_{\odot})$ &  1.76 $\pm$ 0.09 \\
			$a {\mathrm sin}i (R_{\odot})$     & 19.96 $\pm$ 0.36 \\
			\hline 
    \label{tab:orbit}
		\end{tabular}
	\end{center}
\end{table}

\begin{table}
	\begin{center}
		\caption{Kepler data LC model parameters of V396~Gem}
		\begin{tabular}{lcc}\hline\hline
			Parameter              &  Value & Error\\
			\hline
			$P$ (days)             & 5.4956923 & 0.0022831  \\
			$T_{\rm eff1}(K)$      &  \multicolumn{2}{c}{7000}\\
			$T_{\rm eff2}(K)$      &   6978 & 35   \\
			$q$                    &  \multicolumn{2}{c}{0.990 (fixed)} \\
			$L_{1}/L_{1+2}$ (Kep)  &   0.50 & 0.01 \\
			$e$                    &  \multicolumn{2}{c}{0.0 (fixed)}  \\
			$i$ $(^{o})$             &   83.27 & 0.02   \\
			$\Omega_{\rm 1}$       &   8.577 & 0.091 \\
			$\Omega_{\rm 2}$       &   8.552 & 0.118  \\
			${r}_{\rm 1}$      &   0.131 & 0.002 \\
			${r}_{\rm 2}$      &   0.133 & 0.002 \\
			$\chi^{2}$         &  \multicolumn{2}{c}{0.004} \\
			\hline 
	    \label{tab:LCparameter}
		\end{tabular}
	\end{center}
\end{table}

The temperature and metallicity of each component were obtained from the observed spectra fitted by synthetic spectra, which were calculated using {\sc atlas9} \citep[][]{Kurucz1979, Castelli} and {\sc spectrum} \citep[][]{Gray} codes. By using the component light contributions, synthetic spectrum of each component has been combined to have a composite spectrum of the system. H$_{\beta}$  region is selected for temperature determination. We constructed several synthetic spectra with an interval of 100K and noted the $\chi^2$ for each model. After calculating several synthetic spectra, the best model was adopted with temperature of components as 7000 $\pm$ 100 K.

With the final temperatures of the components, the synthetic spectra with different metallicities changing between 0.0 to $+$1.0 dex with an interval of $\Delta[M/H]$ = 0.1 dex were calculated. Eight regions have been selected for determining the metallicity. For each spectral region, metallicity values corresponding to minimum ${\chi}^{2}$ were determined. Then, we have calculated the weighted average of the metallicity ($\Bar{X}_w$) in Eq.~\eqref{average} by adopting weights as $1/{\chi}^{2}$ for each model. In Eq.~\eqref{average}, $X$ refers to the metallicity value for each region, while $w$ and $n$ refer to the weight and number of regions, respectively.

\begin{equation}
    \Bar{X}_w=\frac{\displaystyle\sum_{i=1} ^{n} X_i w_i}{\displaystyle\sum_{i=1} ^{n} w_i}
    \label{average}
\end{equation}

 The weighted standard deviation was calculated by using Eq.~\eqref{deviation} where $\sigma_w$ is the weighted standard deviation, $X_i$ is the determined metallicity for each region, $\Bar{X}_w$ is the determined weighted metallicity, $w$ is the weight for each region, and $n$ is the number of regions.

\begin{equation}
    \sigma_w=\sqrt{\frac{\displaystyle\sum_{i=1} ^{n} w_i \left(X_i-\Bar{X}_w\right)^2}{\frac{\left(n-1\right)\displaystyle\sum_{i=1} ^{n} w_i}{n}}}
    \label{deviation}
\end{equation}

\begin{figure*}
    \centering
    \begin{subfigure}
        \centering
        \includegraphics[width=0.35\textwidth]{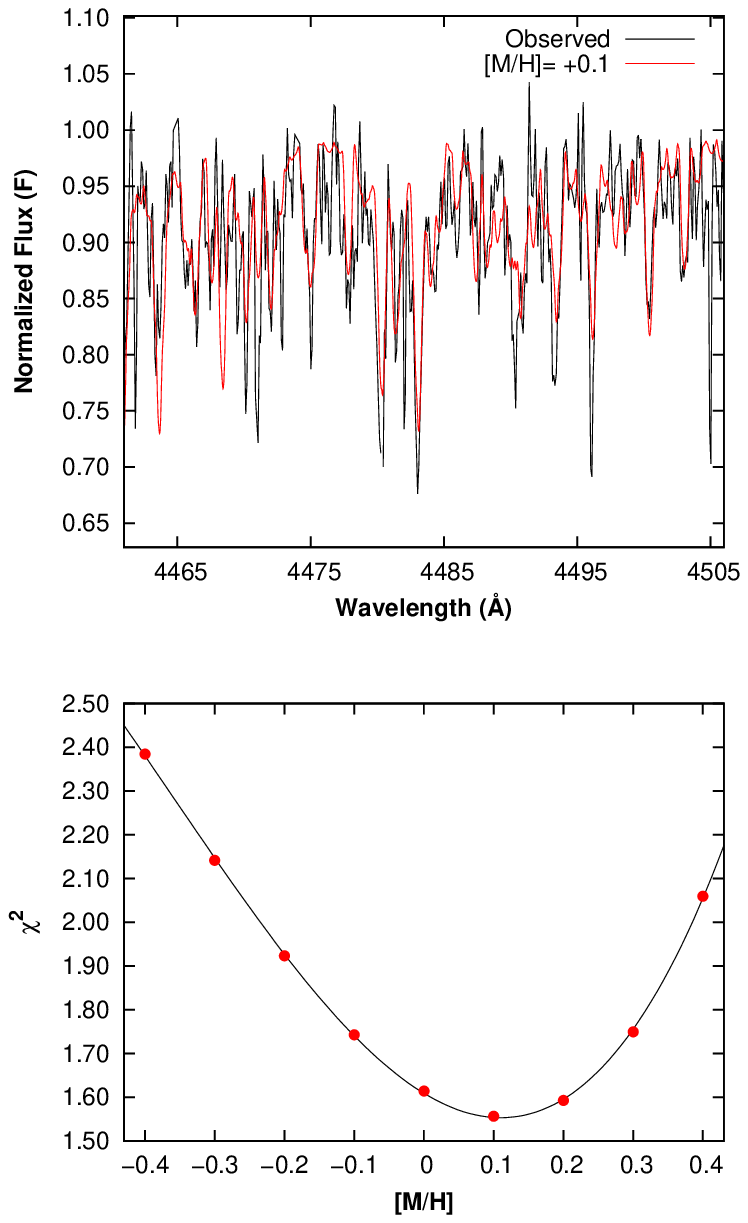}
    \end{subfigure}
    \begin{subfigure}
        \centering
        \includegraphics[width=0.35\textwidth]{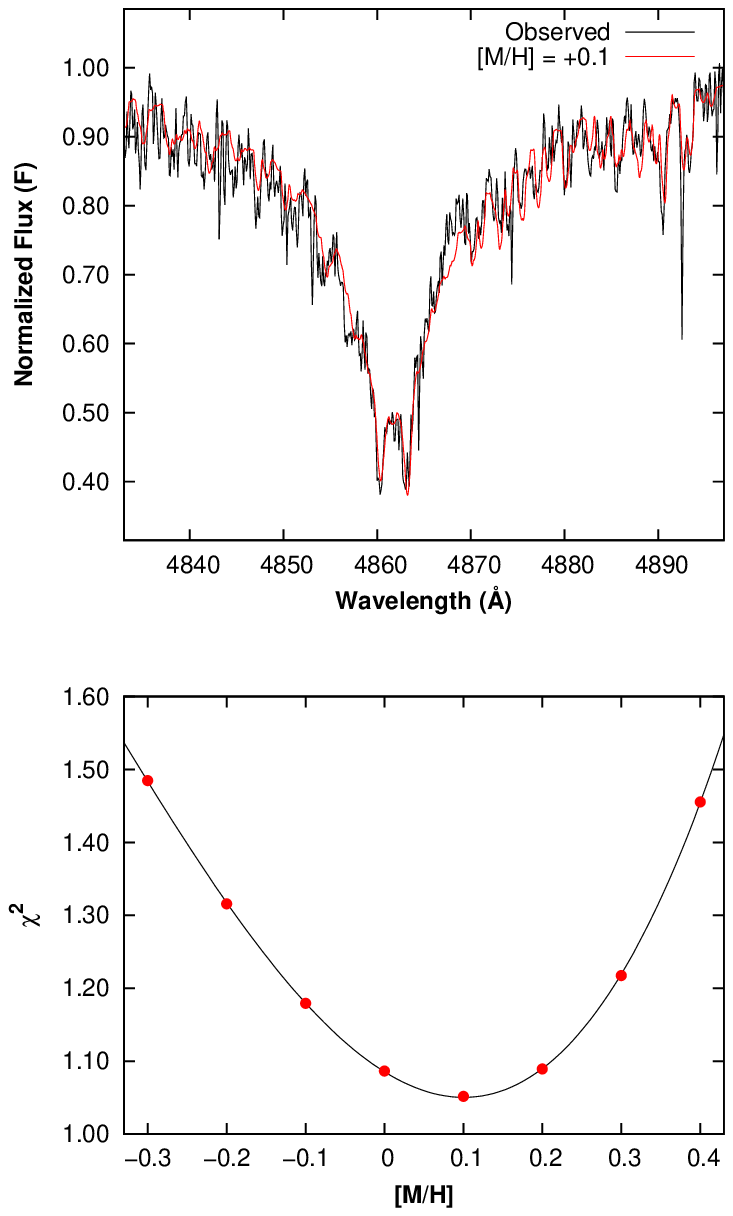}
    \end{subfigure}
    \begin{subfigure}
        \centering
        \includegraphics[width=0.35\textwidth]{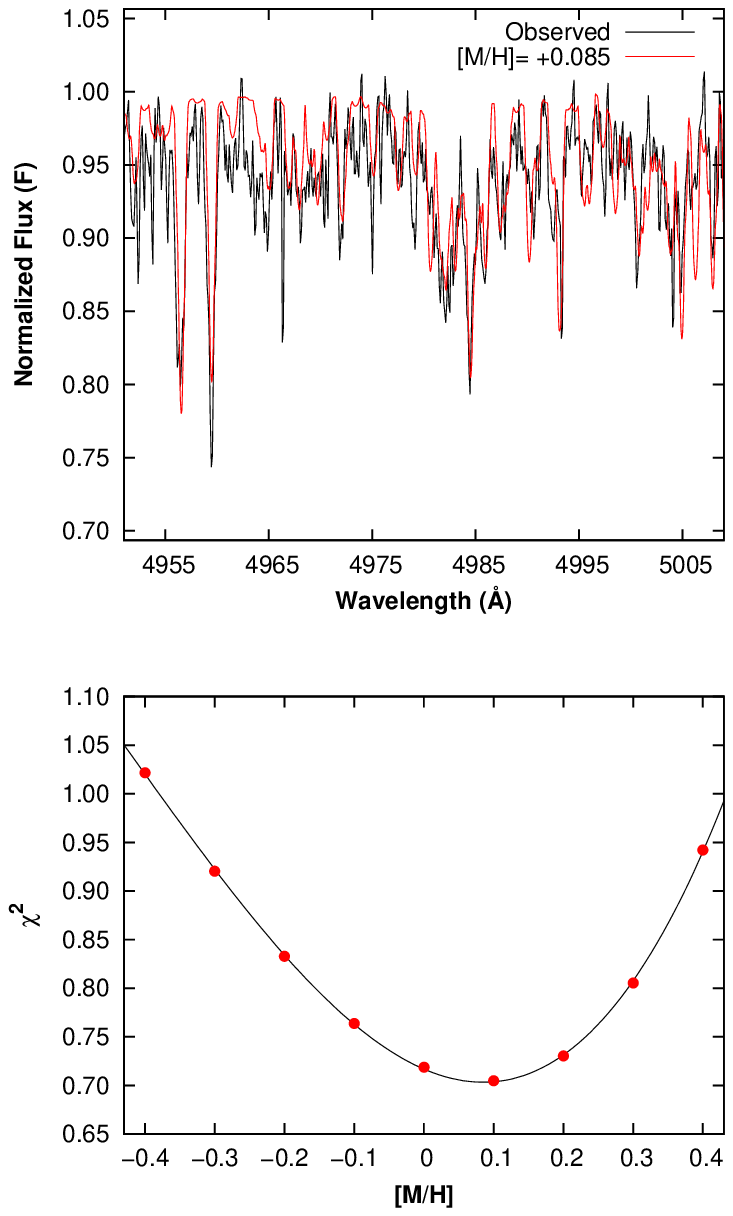}
    \end{subfigure}
    \begin{subfigure}
        \centering
        \includegraphics[width=0.35\textwidth]{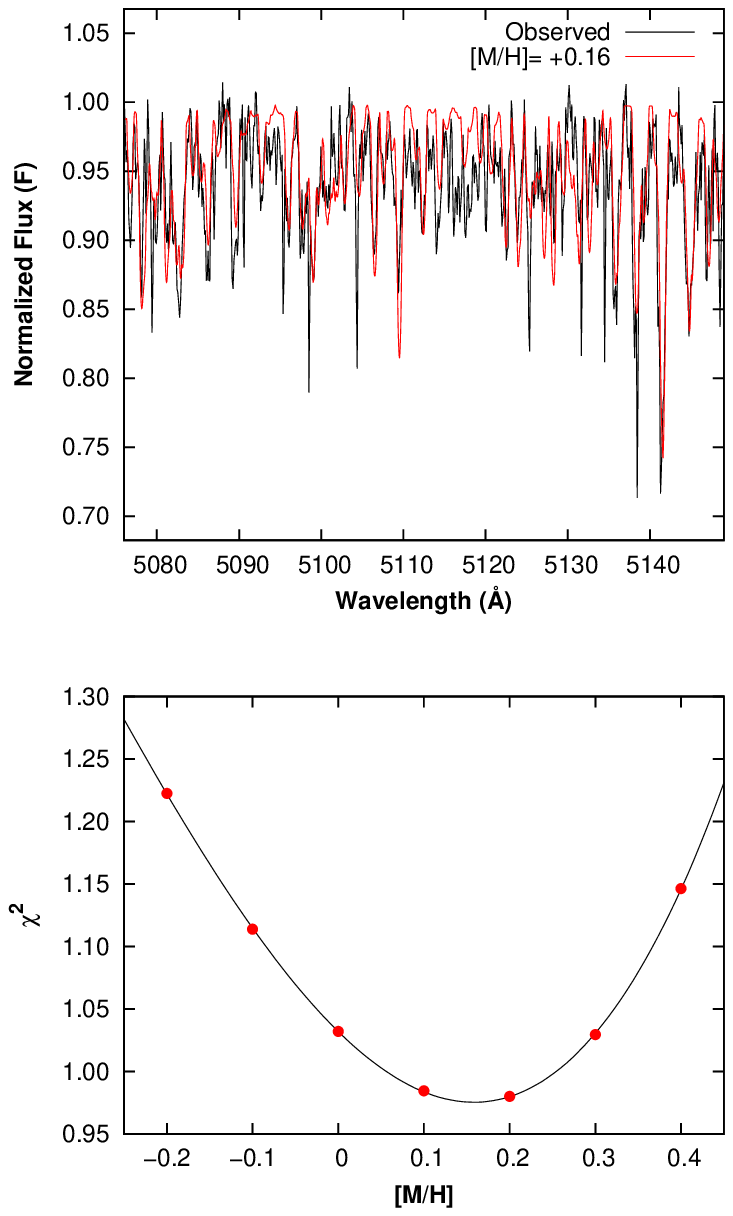}
    \end{subfigure}
    \caption{Some of the selected spectral regions for metallicity determination and ${\chi}^{2}$ fits.}
    \label{fig:orders}
\end{figure*}

Finally, we have determined the metallicity of the system as $[M/H] =~$0.11 $\pm$ 0.03 dex. Selected regions and models for each spectral region are shown in Fig.~\ref{fig:orders}. Astrophysical parameters of V396~Gem have been calculated and are given in Table~\ref{tab:absolutepar}.

\begin{table*}
\small
\caption{Binary stellar parameters of V396~Gem. Errors of parameters are given in parenthesis.} \label{tab:absolutepar}
\begin{tabular}{lccc}\hline
Parameter & Symbol  & Primary & Secondary \\
\hline
Spectral type & Sp & F1 V-IV & F1 V-IV \\
Mass (M$_\odot$) & \emph{M} & 1.814(0.114) & 1.797(0.114) \\
Radius (R$_\odot$) & \emph{R} & 2.655(0.078) & 2.659(0.090)\\
Separation (R$_\odot$) & \emph{a} & \multicolumn{2}{c}{20.098(0.361)} \\
Orbital period (days) & \emph{P} & \multicolumn{2}{c}{5.4956923(0.0022831)} \\
Orbital inclination ($^{\circ}$) & \emph{i} & \multicolumn{2}{c}{83.287(0.020)} \\
Mass ratio & \emph{q} & \multicolumn{2}{c}{0.990(0.036)} \\
Eccentricity & \emph{e} & \multicolumn{2}{c}{0.0(fixed)} \\
Surface gravity (cgs) & $\log g$ & 3.849(0.080) & 3.843(0.085) \\
Metallicity (dex) & [M/H] & \multicolumn{2}{c}{0.11(0.03)} \\
Combined visual magnitude (mag) & \emph{V} & \multicolumn{2}{c}{10.18} \\
Individual visual magnitudes (mag) & \emph{V$_{1,2}$} & 10.71(0.13) &
10.75(0.13)  \\
Combined colour index (mag) & $B-V$ & \multicolumn{2}{c}{0.33(0.02)} \\
Temperature (K) & $T_{\rm eff}$ & 7000(100) & 6978(100) \\
Luminosity (L$_\odot$) & $\log$ \emph{L} & 1.184(0.050) & 1.180(0.054) \\
Bolometric magnitude (mag) & $M_{\rm bol}$ & 1.790(0.125) & 1.800(0.135) \\
Absolute visual magnitude (mag) & $M_{\rm v}$  & 1.759(0.123) & 1.770(0.135) \\
Bolometric correction (mag) & \emph{BC} & 0.031(0.002) & 0.030(0.003) \\
Velocity amplitudes (km\,s$^{-1}$) & $K_{\rm 1,2}$ & 91.46(1.71) & 92.36(1.62) \\
Systemic velocity (km\,s$^{-1}$) & $V_{\gamma}$ & \multicolumn{2}{c}{-3.47(1.24)} \\
Computed synchronization velocities (km\,s$^{-1}$)& V$_{\rm synch}$ & 24.4(0.7) & 24.5(0.8) \\
Observed rotational velocities (km\,s$^{-1}$) & V$_{\rm rot}$ & 25(5) & 25(5)\\
Age (Byrs) & \emph{t} & \multicolumn{2}{c}{1.168(0.149)} \\
Distance (pc) & \emph{d} & \multicolumn{2}{c}{626(72)} \\
GAIA Distance (pc) & \emph{d} & \multicolumn{2}{c}{634(6)}\\
\hline
\end{tabular}
\end{table*}

\subsection{Evolutionary Status of V396~Gem}

\begin{figure}
    \centering
    \begin{subfigure}
        \centering
        \includegraphics[width=0.8\columnwidth]{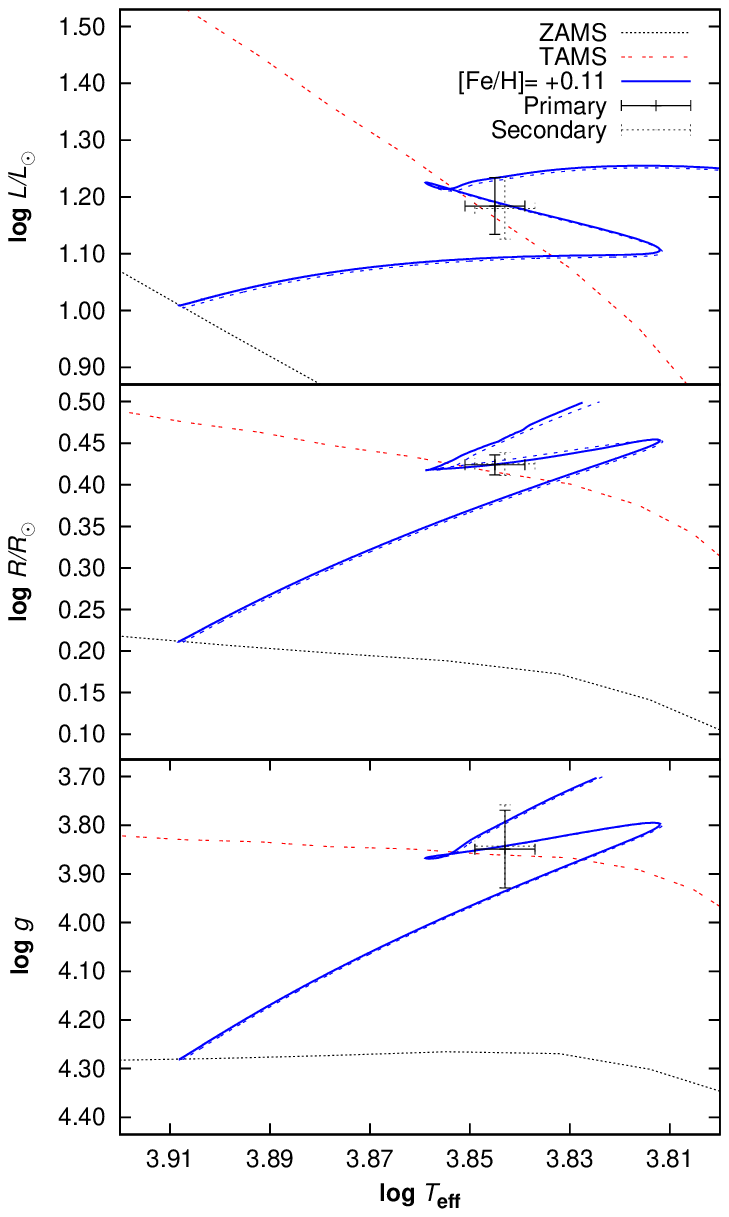}
    \end{subfigure}
    \caption{Location of components of V396~Gem in planes of log $T_{\rm eff}$ - log $L$, log $T_{\rm eff}$ - log $R$, and log $T_{\rm eff}$ - log  $g$ in upper, middle, and lower panels, respectively.}
    \label{fig:evo}
\end{figure}

\begin{figure}
\begin{center}
\includegraphics[width=80mm]{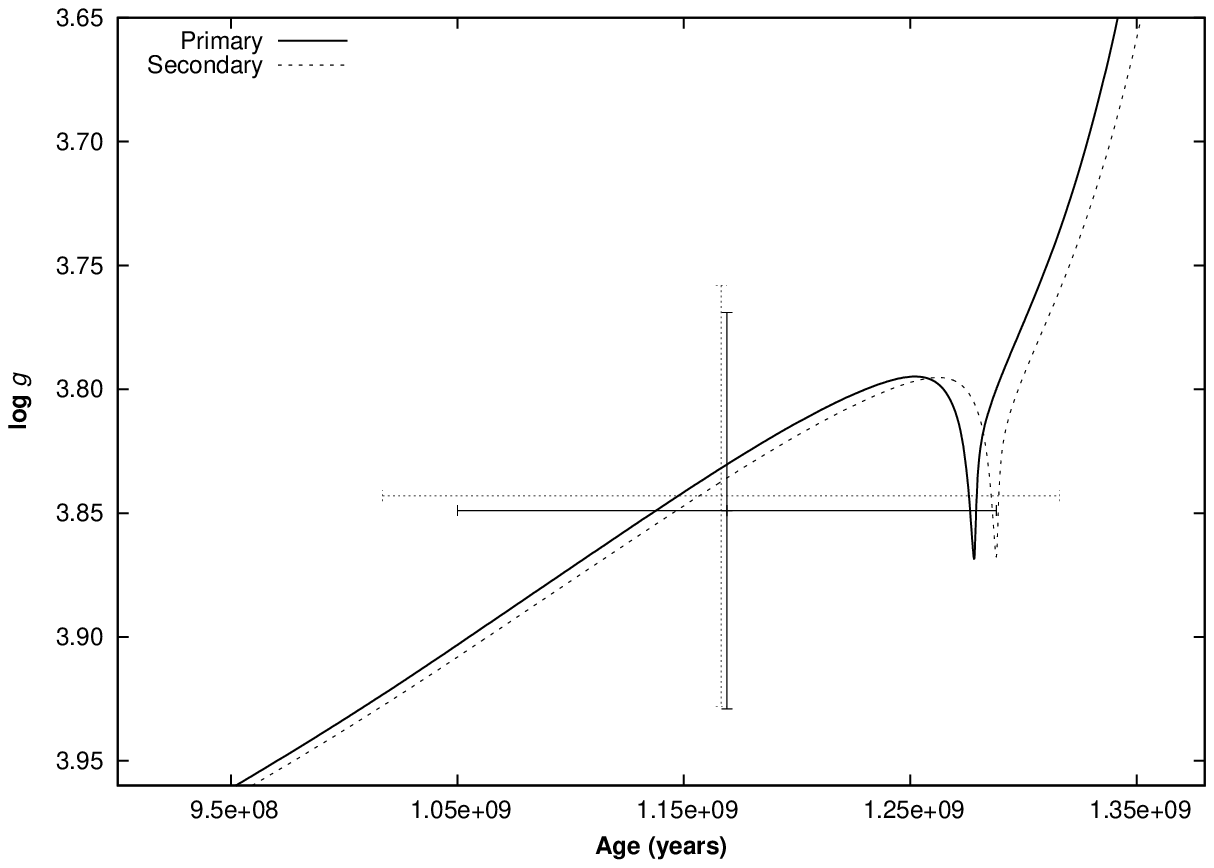}
\caption{Location of the components of V396~Gem in the age-radius plane.}
\label{fig:age}
\end{center}
\end{figure}

We have used one-dimensional stellar evolution code  MESA v12115 (Modules for Experiments in Stellar Astrophysics; \cite{Paxton2011, Paxton2013, Paxton2015, Paxton2018, Paxton2019}) to analyze the evolutionary status of V396~Gem. A rotating single star evolution scenario for both stars is adopted. Since the components are detached, evolutionary lines with $[M/H] = +0.11$ dex were calculated for each component. According to the evolutionary status shown in Fig.~\ref{fig:evo}, the components of V396~Gem are still on the main-sequence, burning hydrogen in their cores and getting closer to the terminal-age main sequence (TAMS). As it can be seen in Fig.~\ref{fig:evo}, components of V396~Gem is in a unique place where parameters change in a short time span, which allows us to determine the age of the system more reliable. The age - log $g$ diagram shown in Fig.~\ref{fig:age} is a useful tool to determine the age and its uncertainty for the system since it includes both mass and radius information of the components are obtained with good precision ($\sim$6 per cent) (see Table\,\ref{tab:absolutepar}). According to the locations of the components in Fig.~\ref{fig:age}, the age of the system is 1.168$\pm$0.149 billion years.

\section{Discussion and Conclusions}
\label{sec:discussion}

\begin{figure}
    \begin{center}
        \includegraphics[width=85mm]{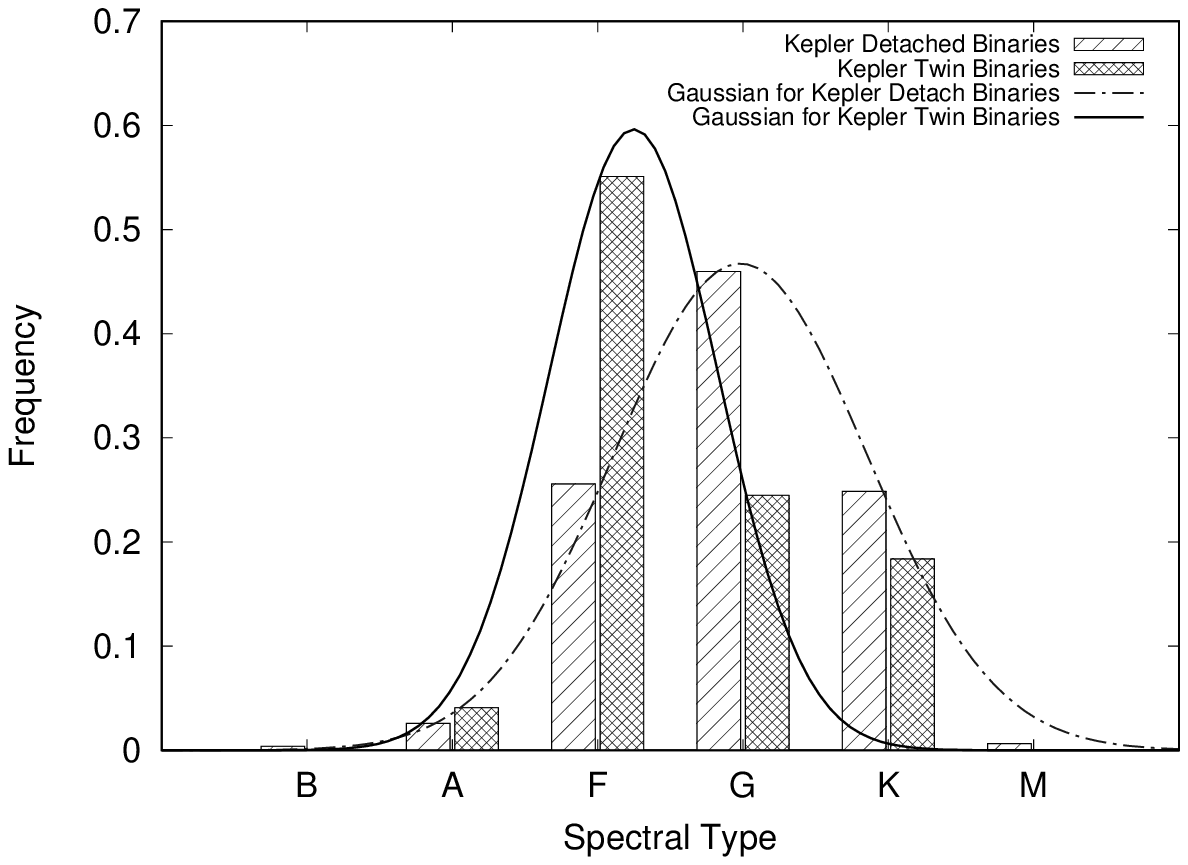}
        \caption{Distribution of spectral types between Kepler detached binaries and Kepler twin binaries.}
    \label{fig:spec_dist_kepler}
    \end{center}
\end{figure}

We have identified eclipsing twin binary systems observed by the Kepler telescope. According to our classification, among 2920 eclipsing binary systems in KEBC, 49 systems (34 circular, 15 eccentric orbit) are determined as highly probable twin binary. Spectral types of circular orbit twins have been determined with new photometric observations and of eccentric orbit twins with temperature data given by \cite{Pinson2012}. 
In Fig.~\ref{fig:spec_dist_kepler}, spectral type distribution of Kepler twins are compared with Kepler detached binaries. The tendency of twin binaries among the Kepler detached binaries is toward F-spectral type with a Gaussian peak at F2-3, where Kepler detached binaries favour G-spectral type with a Gaussian peak at G-type, which supports \cite{bakis2020} who studied all sky automated survey (ASAS) \citep{pojmanski} database for twins.

\begin{figure}
    \begin{center}
        \includegraphics[width=85mm]{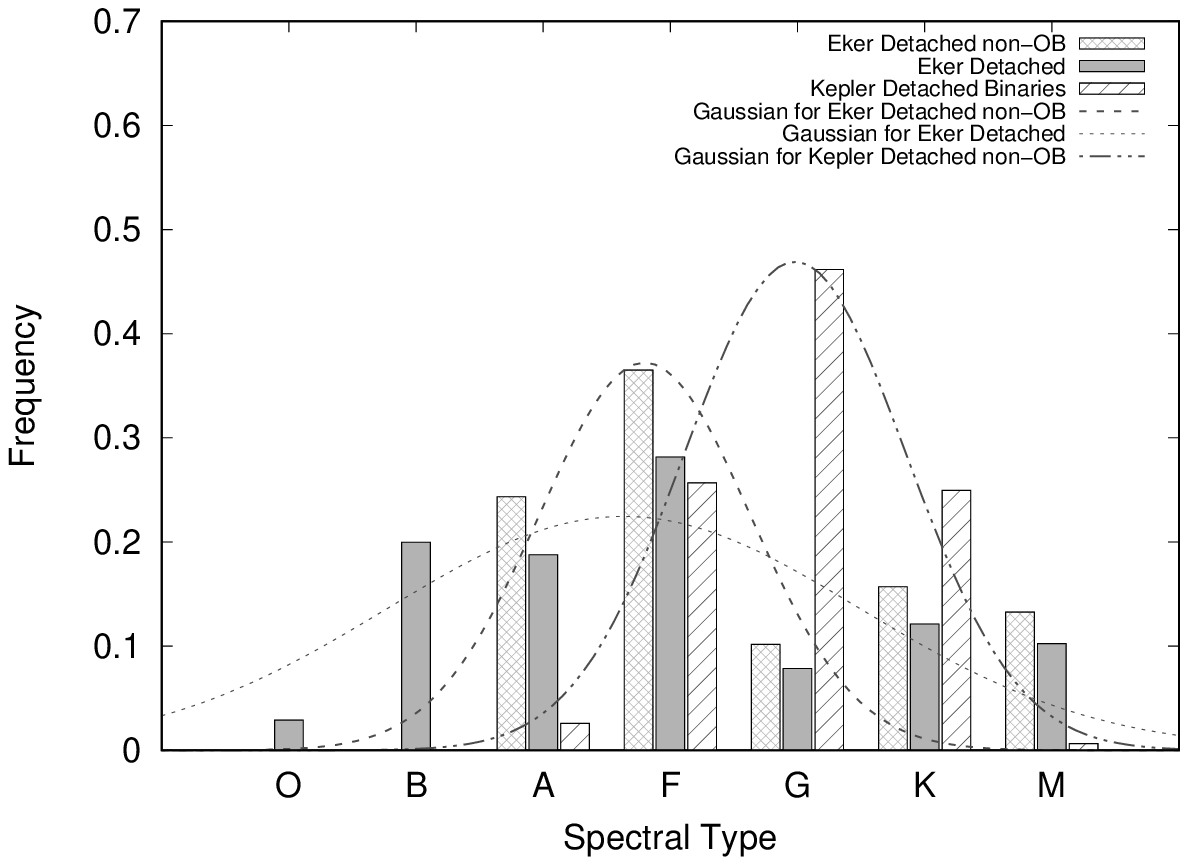}
        \caption{Spectral type distribution of detached binaries in two samples: \citet{Eker} without OB binaries and KEBC.}
    \label{fig:spec_dist_no_OB}
    \end{center}
\end{figure}

\begin{figure}
    \begin{center}
        \includegraphics[width=85mm]{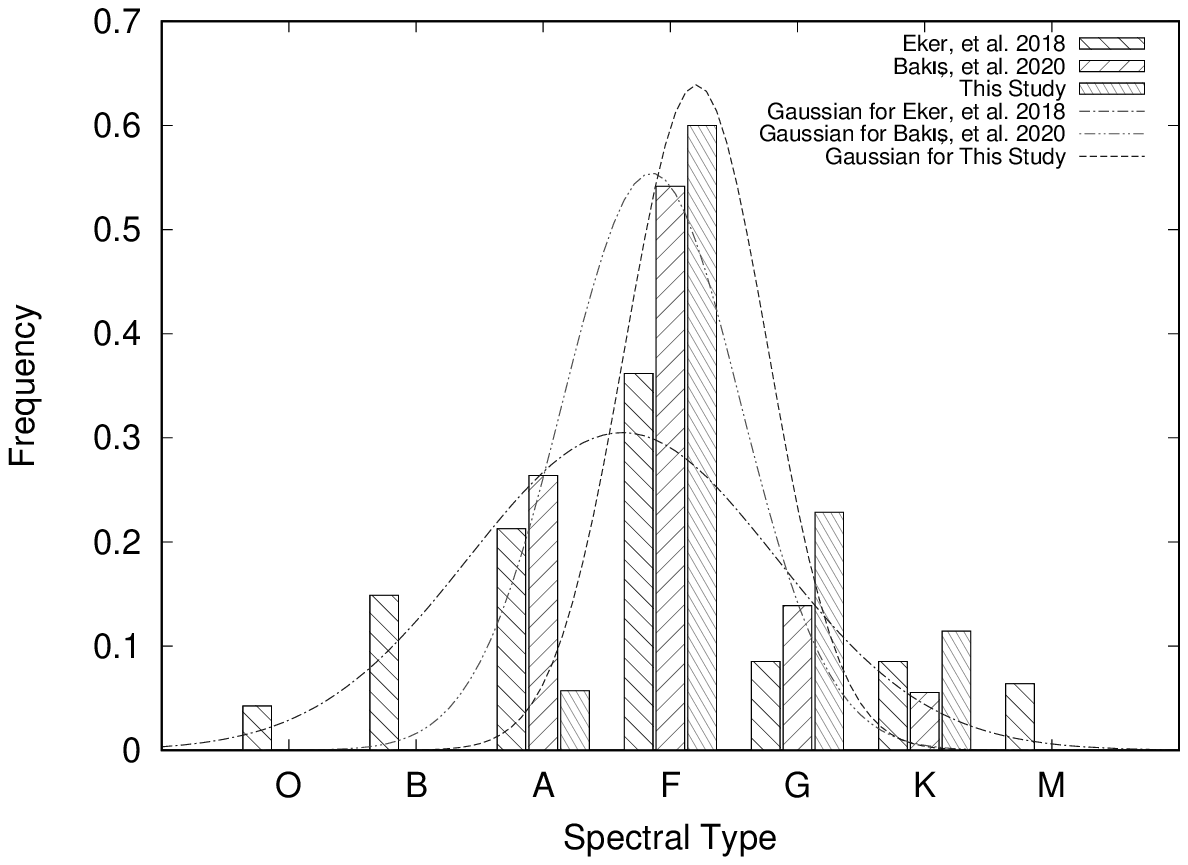}
        \caption{Distribution of spectral types of all known eclipsing twin binaries.}
    \label{fig:spec_dist_twin}
    \end{center}
\end{figure}

The Kepler telescope observed late spectral type stars. Therefore, sample selection bias is expected when comparing Kepler data with the spectral type independent sample of \cite{Eker}. In order to see the relation of two samples, we removed the early type stars from the sample of \cite{Eker} and compared them both samples in Fig.\ref{fig:spec_dist_no_OB}.  There is nearly one spectral type difference between the two samples, which may explain the shift of Kepler twin binaries to later spectral types among all known twin binaries in Fig.\ref{fig:spec_dist_twin}. As the distribution of twins is similar to the one of the detached binaries, this study agrees with the idea that the twin binaries do not have a spectral type preference among the detached eclipsing binary systems.

\begin{figure}
    \begin{center}
        \includegraphics[width=85mm]{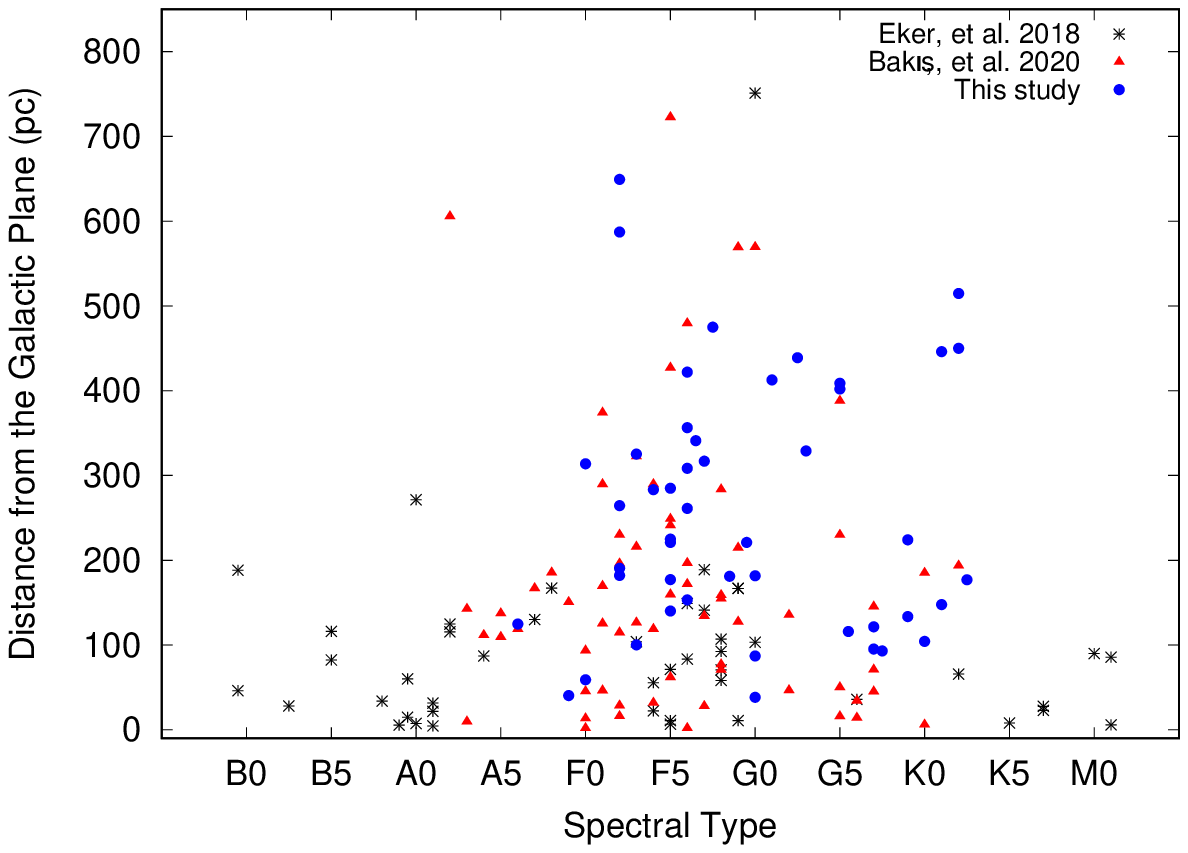}
        \caption{Distance of eclipsing twin binary systems from the galactic plane.}
    \label{fig:galactic}
    \end{center}
\end{figure}

The distance of the twin binaries to the galactic plane is shown in Fig.~\ref{fig:galactic}. In the figure, while all twins are located in the galactic disk, early (B and A) and late (K and M) spectral type twins are not located far from the galactic plane as F and G-type systems are. It is expected that early-type systems are located in the spiral arms where they are formed. Although, M-type stars are found all over the galaxy, relatively less abundance of late-type twins is a question needs to be addressed. Nevertheless, there is no explanation why F-type twins are spread over more distances from the galactic plane than the other spectral type twins are. The most reasonable explanation at the moment seems to be the selection effect as F-type systems are the most abundant among twins. Additionally, from the location of twins in the galactic disk, we can say that twins may belong to the galactic thin disk population.

Among our final list of twin binaries, V396~Gem (EPIC 202072991) was selected to analyze with spectroscopic data to obtain the fundamental parameters. This system is an early F-type of, which twins are most abundant. The analysis of spectral data over the visible region enabled us to determine the metal abundance of the system as $[M/H]=0.11\pm0.03$ dex, which is a good accuracy compared to similar studies in the literature. As can be seen from Figure \ref{fig:evo}, the V396 Gem is between the main-sequence turn-off point and TAMS. However, our uncertainty box in the log $T_{\rm eff}$ - log $g$ plane is wider than the others, which is adopted as the uncertainty for our age estimation. In Figure \ref{fig:age}, we show the relationship between age and log $g$.

It should be noted that our twin binary list does not cover all eclipsing twin binary systems that were observed by the Kepler telescope. Some of the eclipsing twin binary systems, \citep[eg. KIC 9402652,][]{Koo}, are not in our list since they do not meet our selection criteria or their components are evolved \citep[eg. KIC 10031808,][]{Helminiak2019}. Recently, \cite{Bulut} has studied 4 twin binary systems with combined spectroscopic and photometric data. They state that the twin candidates they studied were selected by their visual inspection. However, as their selection criteria are not within the our limits, our list excludes those twin binaries (EPIC 202843107, EPIC 204321014, EPIC 216075815, and EPIC 217988332).

\section*{Acknowledgements}

We are grateful to the anonymous referee for her/his valuable suggestions. This work is part of the PhD thesis of GY. We thank to T\"{U}B{\.{I}}TAK for funding this research under project number 115F029, to TUG for a partial support in using T100 telescope with project number 20AT100-1657 and to Akdeniz University Space Sciences and Technologies Department for granting us observing time with UBT60 telescope. This paper includes data collected by the $Kepler$ mission. Funding for the $Kepler$ mission is provided by the NASA Science Mission Directorate. This research has made use of the SIMBAD database, operated at CDS, Strasbourg, France. This work has made use of data from the European Space Agency (ESA) mission {\it Gaia} (\url{https://www.cosmos.esa.int/gaia}), processed by the {\it Gaia} Data Processing and Analysis Consortium (DPAC,
\url{https://www.cosmos.esa.int/web/gaia/dpac/consortium}). Funding for the DPAC has been provided by national institutions, in particular the institutions participating in the {\it Gaia} Multilateral Agreement.

\section*{Data Availability}

The data underlying this article are available in the article and in its online supplementary material.




\bibliographystyle{mnras}
\bibliography{reference} 








\bsp	
\label{lastpage}
\end{document}